\documentclass[10pt]{article}
\usepackage{amssymb}
\usepackage{geometry}

\usepackage{graphicx}
\chardef\bslash=`\\ 

\begin{document}

\title{Re-determination of the Pseudobinary System Li$_2$O -- MoO$_3$ \\
\small excerpt for copyright reasons, full article published: \\ Cryst. Res. Technol. 43, No. 4, 350--354 (2008) / DOI 10.1002/crat.200711106}

\author{M. Moser\footnote{Permanent address: Humboldt-Universit\"at zu Berlin, Institut f\"ur Chemie, Brook-Taylor-Str. 2, 12489 Berlin, Germany}, D. Klimm\footnote{Corresponding author: e-mail: {\sf klimm@ikz-berlin.de}}, S. Ganschow, A. Kwasniewski, K. Jacobs \\ Institute for Crystal Growth \\ Max-Born-Str. 2, 12489 Berlin, Germany}

\date{21 December 2007}

\maketitle

\section*{Results and Discussion}

For the components Li$_2$O and MoO$_3$ just the following intermediate phases were confirmed independently by other authors using X-ray diffraction. It must be assumed that the phases Li$_2$Mo$_2$O$_7$ and Li$_2$Mo$_3$O$_{10}$, which are mentioned in Tab.~\ref{tab:phases}, do not exist.
\begin{itemize}
	\item Li$_4$MoO$_5$: LiO$_{0.5}$ : MoO$_3$ = 4.2 : 1.0 were annealed at $950^{\,\circ}$C for 28 days. Crystals are triclinic (disordered rocksalt type, $P\,\bar{1}$) with $a_0=5.1094(5)$\,\AA, $b_0=7.7169(7)$\,\AA, $c_0=6.0609(4)$\,\AA, $\alpha=101.804(8)^\circ$, $\beta=101.78(1)^\circ$, $\gamma=108.770(9)^\circ$ \cite{Hoffmann89}.
	\item Li$_2$MoO$_4$: Single crystals can be obtained \cite{Brower72}, belonging to the phenacite type structure type ($R\,\bar{3}$ with $a_0=14.330(2)$\,\AA, $c_0=9.584(2)$\,\AA) \cite{Kolitsch01}. The substance crystallizes in anhydrous form from aqueous solution \cite{Hoermann29}.
	\item Li$_4$Mo$_5$O$_{17}$: Single crystals of about 1\,cm diameter can be grown by the Czochralski technique \cite{Brower72} ($P\,\bar{1}$ with $a_0=6.7775(1)$\,\AA, $b_0=9.461(1)$\,\AA, $c_0=10.802(2)$\,\AA, $\alpha=73.16(2)^\circ$, $\beta=88.98(1)^\circ$, $\gamma=69.760(2)^\circ$ \cite{Wiesmann97}).
	\item Li$_2$Mo$_4$O$_{13}$: $40\times50\times60$\,$\mu$m$^3$ sized crystals of the low-$T$ form ($P\,\bar{1}$ with $a_0=8.578(5)$\,\AA, $b_0=11.450(5)$\,\AA, $c_0=8.225(5)$\,\AA, $\alpha=109.24(7)^\circ$, $\beta=96.04(7)^\circ$, $\gamma=95.95(7)^\circ$) were obtained by slow cooling of Li$_2$CO$_3$/MoO$_3$ melts \cite{Gatehouse74}. $30\times30\times60$\,$\mu$m$^3$ sized crystals of the high-$T$ form ($P\,\bar{1}$ with $a_0=8.612(4)$\,\AA, $b_0=11.562(6)$\,\AA, $c_0=8.213(4)$\,\AA, $\alpha=94.45(7)^\circ$, $\beta=96.38(7)^\circ$, $\gamma=111.24(7)^\circ$) were prepared by annealing mixtures of Li$_2$MoO$_4$ and MoO$_3$ for 18~hours at $547^{\,\circ}$C. Besides, a monoclinic form was reported ($P\,2_1/m$ or $P\,2_1$) \cite{Gatehouse75}. 
\end{itemize}

\begin{table}
\caption{Crystalline Li--Mo$^\mathrm{VI}$--O phases as reported by different authors. $x$ is the molar fraction of MoO$_3$. ``i'' and ``c'' means incongruent or congruent melting, respectively, as reported in the given references.}
\renewcommand{\arraystretch}{1.5}
\begin{tabular}{lrrrrr} \hline
Formula      &  $x$                              &$T_\mathrm{f}$ ($^\circ$C)& melting & references               & this study \\
\hline
Li$_2$O=LiO$_{0.5}$ & 0.000                      & $1570$                   & c             & \cite{FactSage5_5} & \\
Li$_4$MoO$_5$       & 0.200                      & $>950$                   &               & \cite{Hoffmann89,Blasse64} & i?\\
Li$_2$MoO$_4$       & 0.333                      & $705$                    & ?                     & \cite{Hoermann29} & $T_\mathrm{f}=698^{\,\circ}$C (i) \\
Li$_2$Mo$_2$O$_7$   & 0.500                      & $530-535$                & i/c & \cite{Hoermann29,Parmentier72}      & not found \\
Li$_4$Mo$_5$O$_{17}$& 0.556                      & $544$                    & c             & \cite{Brower72}           & $T_\mathrm{f}=547^{\,\circ}$C (i) \\
Li$_2$Mo$_3$O$_{10}$& 0.600                      & $549$                    & i          & \cite{Hoermann29}            & not found \\
Li$_2$Mo$_4$O$_{13}$& 0.667                      & $568-576$                & i           & \cite{Hoermann29,Brower72,Parmentier72} & $T_\mathrm{f}=570^{\,\circ}$C (i) \\
MoO$_3$             & 1.000                      & $795-801$                & c           & \cite{Hoermann29,Brower72,FactSage5_5,Lisnyak00,Cosgrove53} & $T_\mathrm{f}=782^{\,\circ}$C (c) \\
\hline
\end{tabular}
\label{tab:phases}
\end{table}

\begin{figure}
\includegraphics[width=.48\textwidth]{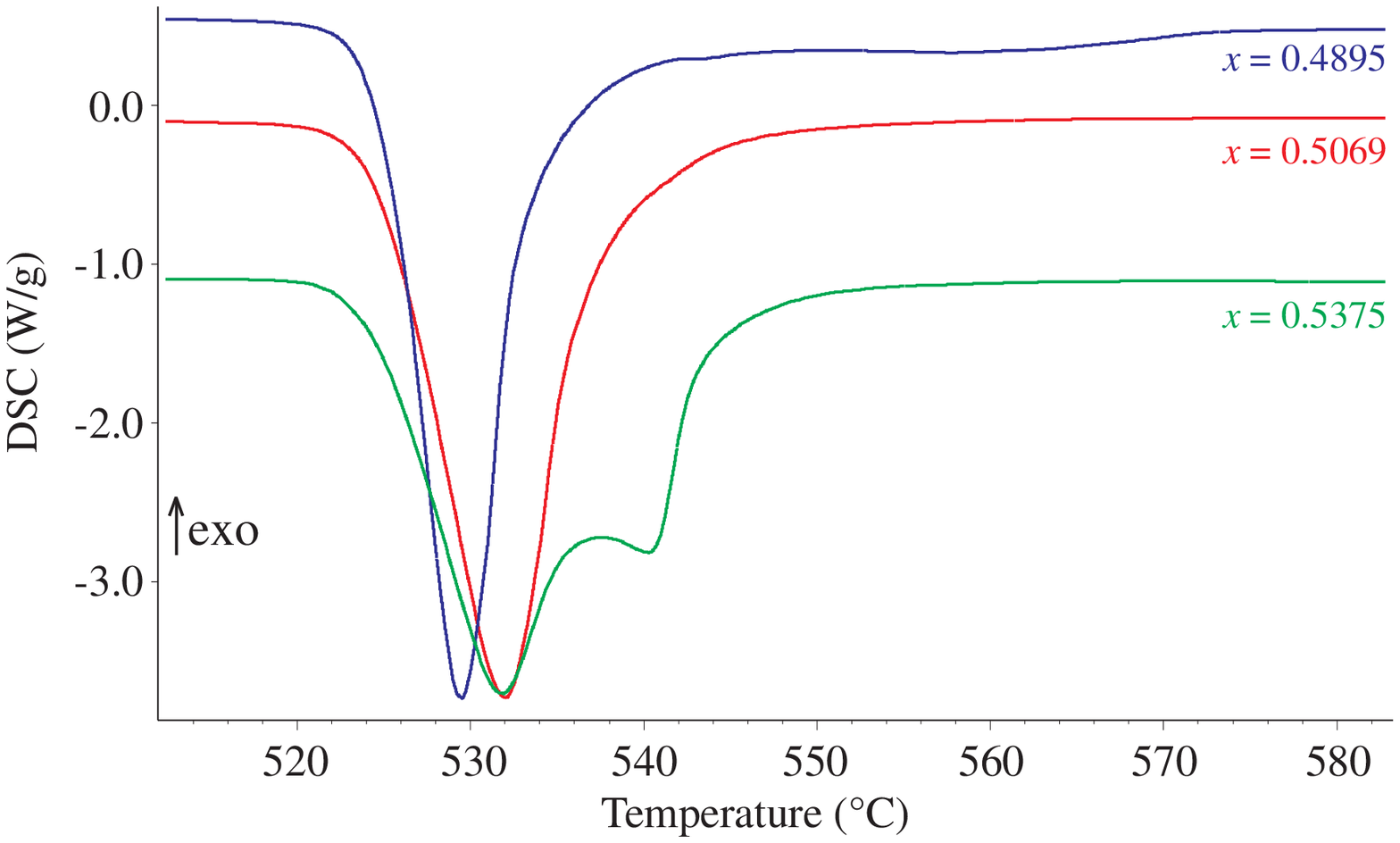}~a)
\hfil
\includegraphics[width=.42\textwidth]{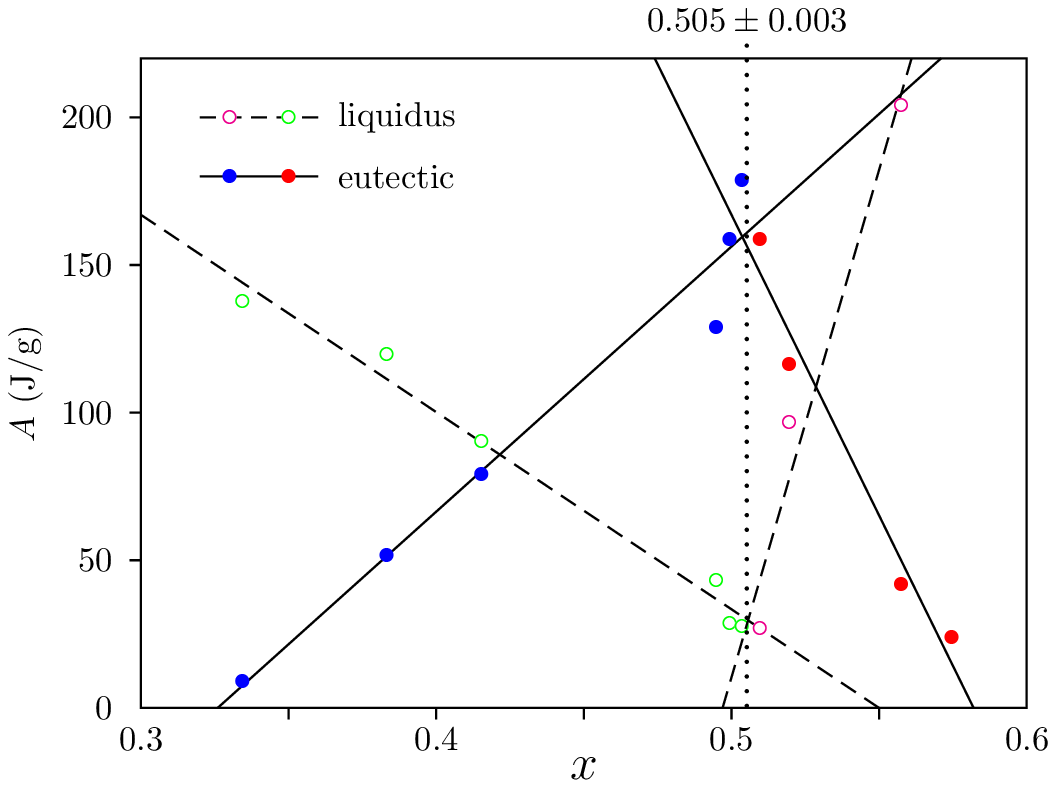}~b)
\caption{\newline a) From top to bottom: DSC heating curves on the Li$_2$O-rich side, in the vicinity of the eutectic composition $x_\mathrm{eut}=0.505$, and on the MoO$_3$-rich side. \newline b) Peak areas $A(x)$ for the eutectic peak and for the shoulder extending to the liquidus.}
\label{fig:DSC}
\end{figure}

\begin{figure}
\includegraphics[width=.7\textwidth]{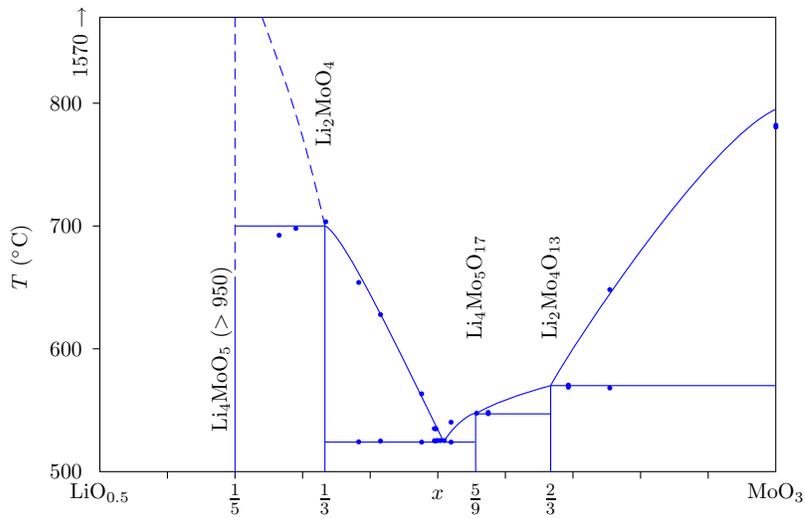}
\caption{The phase diagram Li$_2$O-MoO$_3$ with some experimental points, as determined by DSC.}
\label{fig:PD}
\end{figure}

\clearpage


\end{document}